\documentclass[aps,prl,twocolumn,superscriptaddress,nofootinbib,floatfix]{revtex4-2}
\usepackage{amsmath,amssymb}
\usepackage{graphicx}
\usepackage[colorlinks,linkcolor=blue,citecolor=blue,urlcolor=blue]{hyperref}

\newcommand{\shZ}{\hat s_Z}
\newcommand{\QW}{Q_W^e}
\newcommand{\MSbar}{$\overline{\rm MS}$}
\newcommand{\ee}{\times10^{-3}}
\newcommand{\dd}{\text{d}}

\newcommand{\Qref}{0.0056}

\newcommand{\Btree}{75.12}

\newcommand{\BoneB}{3.01}
\newcommand{\BoneF}{-33.34}
\newcommand{\BoneFL}{-33.64}
\newcommand{\BoneTU}{0.05}
\newcommand{\BoneAS}{-0.08}
\newcommand{\BtwoB}{-0.45}
\newcommand{\BtwoF}{1.39}
\newcommand{\BtwoFL}{1.39}
\newcommand{\BtwoFF}{-0.26}
\newcommand{\BtwoFFL}{-0.27}
\newcommand{\BtwoAS}{0.02}

\newcommand{\BtwoSum}{0.70}

\newcommand{\Bdata}{45.46}
\newcommand{\Blat}{45.17}
\newcommand{\Edata}{0.07}
\newcommand{\Elat}{0.02}
\newcommand{\Bdiff}{0.29}

\newcommand{\ALRref}{40.9}

\newcommand{\Bdl}{0.12}
\newcommand{\Lref}{13.5}
\newcommand{\dsGoal}{0.00032}

\newcommand{\dBdE}{0.022}
\newcommand{\BmeasF}{0.043}

\newcommand{\Bpert}{0.04}

\newcommand{\BtwoPct}{1.5}
\newcommand{\BdiffPct}{0.6}
\newcommand{\BThrOne}{-3.49}
\newcommand{\BThrTwo}{0.10}

\begin{document}

\title{Parity-violating M\o ller scattering: complete two-loop corrections}

\author{Andrzej Czarnecki}
\affiliation{Department of Physics, University of Alberta, Edmonton, Alberta, Canada T6G 2E1}

\begin{abstract}
The MOLLER experiment will measure the parity-violating asymmetry of M\o ller scattering to 2.4\%. We present its fermionic and bosonic two-loop electroweak corrections to the zeroth and first orders in $1-4\sin^2\theta_W$. 
Light quarks enter through vacuum polarization from  $e^+e^-$ data or lattice QCD.
For $\sin^2\hat\theta_W(M_Z) = 0.23122$ and an 11 GeV beam, the weak charge of the electron is $0.04546(8)$ (data) or 0.04517(5) (lattice). Two-loop terms contribute 1.5\% of it. Uncomputed terms are estimated at 4\% of the experimental goal; the data-lattice difference, about 25\% of that goal, limits the prediction.
\end{abstract}

\maketitle

Left- and right-polarized electrons scatter from unpolarized electrons with slightly different cross sections $\sigma_{L,R}$. This parity-violating (PV) asymmetry,
$A_{LR}=(\dd\sigma_L-\dd\sigma_R)/(\dd\sigma_L+\dd\sigma_R)$, results from the interference of photon $\gamma$ and the neutral boson $Z$ exchanges. The $Z$ couples to the vector current of one electron and to the axial current of the other~\cite{Derman:1979zc}. 

The vector coupling of the $Z$ to electrons is proportional to $1-4\sin^2\theta_W$,
where $\theta_W$ is the weak mixing angle. 
Since $\sin^2\theta_W\approx0.23$ is close to
$1/4$, the asymmetry is small,
and thus a sensitive probe of new physics: a force with unsuppressed coupling to electrons need only be feeble to shift $A_{LR}$ noticeably~\cite{Kumar:2013yoa}.

Radiative corrections  (RC) can also change $A_{LR}$ significantly because they include PV diagrams without the vector coupling of $Z$, notably the $\gamma Z$ mixing. Indeed, one-loop RC reduce $A_{LR}$ by about 40\%~\cite{Czarnecki:1995fw}. 
This spared the E158 experiment at SLAC~\cite{SLACE158:2005uay,Czarnecki:2005pe} the need to build an expensive new polarimeter because the smaller asymmetry meant it could measure the electron beam polarization less accurately.  Its successor, MOLLER at Jefferson Lab~\cite{MOLLER:2014iki}, aims to measure $A_{LR}$ more precisely by at least a factor of 5. 
This will determine $\theta_W$ about as precisely as the best single
measurements at the $Z$ pole~\cite{ALEPH:2005ab}.

Here we  evaluate  two-loop RC and predict $A_{LR}$ within the Standard Model, to give  MOLLER a benchmark to interpret its result. We write the asymmetry as~\cite{Czarnecki:1995fw}
\begin{equation}
A_{LR}=\frac{G_\mu Q^2}{\sqrt2\,\pi\alpha}\,\frac{1-y}{1+y^4+(1-y)^4}\;\QW(y,s),
\label{eq:alr}
\end{equation}
where $Q^2 = ys$ is the momentum transfer, $s$ is the square of the center-of-mass energy, $\alpha \simeq 1/137$,  $G_\mu$ is the Fermi constant \cite{Marciano:1999ih}, and $\QW$ is the so-called weak charge of the electron. $\QW$ is proportional to the vector coupling of the electron to the $Z$ at tree level but is modified by loops. Normalized to $G_\mu$ as in Eq.~\eqref{eq:alr}, $\QW$ is a function of $\alpha$,  the masses $M_Z$, $M_H$, $m_t$ on shell, and $\shZ \equiv \sin\hat\theta_W(M_Z)$, the sine of the mixing angle with minimal subtraction (\MSbar) as defined in Ref.~\cite{ParticleDataGroup:2026mpi}. In $\shZ$ the logarithm of the top-quark mass is removed from the $\gamma Z$ mixing~\cite{Marciano:1990dp,Fanchiotti:1992tu,Gambino:1993dd}.
Thus
\begin{equation}
\QW=1-4\shZ^2+\frac{\alpha}{\pi}\,b_1(\shZ^2)
+\Big(\frac{\alpha}{\pi}\Big)^2\Big[b_2+(1-4\shZ^2)\,b_2^\prime\Big]+\dots
\label{eq:exp}
\end{equation}
The one-loop coefficient $b_1$ is known for any value of $\shZ$~\cite{Czarnecki:1995fw}.
 At two loops we expand around $\shZ^2=1/4$ and keep two terms of the expansion. In diagrams with two fermion loops we keep the exact dependence on $\shZ$. Its second order term, $2\times 10^{-6}$, serves as the uncertainty of the truncation for the other diagrams.  We use $m_e^2\ll s\ll M_Z^2$ and keep the leading term in both
ratios.
\begin{figure}[htb]
\includegraphics{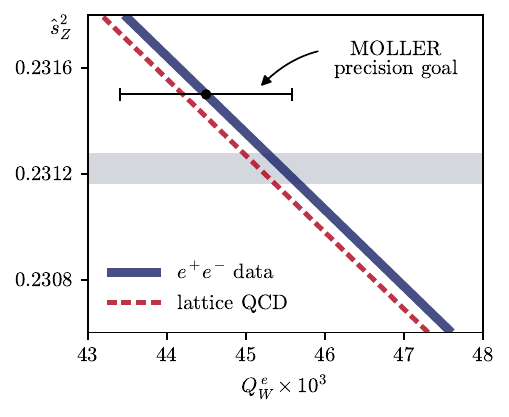}
% fig_qw.py, from qw_s2.dat and numbers.tex
\caption{The relation between $\shZ^2$ and the measured weak charge of the
electron, for MOLLER kinematics, with
the hadronic input from $e^+e^-$ data or from lattice QCD (dashed).
Band widths are the  uncertainties in Eq.~\eqref{eq:main}, added in quadrature.
The grey
band is the Standard Model $\shZ^2$~\cite{ParticleDataGroup:2026mpi}. The bar shows MOLLER's precision goal, $\pm2.4\%$, drawn at an arbitrary point. \label{fig:qw}}
\end{figure}
 For
a beam of $11$~GeV and the mixing angle of the Review of Particle Physics,
$\shZ^2=0.23122$~\cite{ParticleDataGroup:2026mpi}, we find
\begin{equation}
10^3\,\QW=
\begin{cases}
\Bdata\pm\Edata\pm\Bpert & \text{($e^+e^-$ data)},\\
\Blat\pm\Elat\pm\Bpert & \text{(lattice QCD)}.
\end{cases}
\label{eq:main}
\end{equation}
The first uncertainty comes from the hadronic input, the second from terms beyond this
calculation. MOLLER accepts center-of-mass angles from $60^\circ$ to
$120^\circ$~\cite{MOLLER:2014iki}. Across this range the weak charge changes by only
$10^{-5}$, so Eq.~\eqref{eq:main} serves for the whole acceptance. A
measured weak charge $Q_W^{\rm exp}$ determines the mixing angle (both charges in absolute units),
\begin{equation}
\shZ^2=0.23122+\frac{\QW-Q_W^{\rm exp}}{3.45}.
\label{eq:angle}
\end{equation}
The goal of MOLLER, $\pm 1.1 \ee$ in the weak charge, corresponds to $\pm\dsGoal$ in the
angle (Fig.~\ref{fig:qw}).
The ancillary files~\cite{Czarnecki:2026anc} contain all coefficients, a program that
reproduces Tables~\ref{tab:res} and~\ref{tab:cmp}, and the amplitudes of all diagrams, grouped as in their catalogue.

\emph{Calculation.}---The diagrams are generated with \textsc{qgraf}~\cite{Nogueira:1991ex}
and evaluated with \textsc{form}~\cite{Vermaseren:2000nd,Davies:2026cci} in the Feynman
gauge. There are $5326$ two-loop diagrams of M\o ller scattering without closed fermion
loops and $938$ with them, apart from tadpole diagrams, which vanish or cancel. The normalization
to $G_\mu$ requires muon decay at the same
order ($2772$ two-loop diagrams) and the two-loop self-energies of the leptons.

The amplitudes are written in the basis of Ref.~\cite{Glover:2004si}, from which the
one-loop and two-loop amplitudes of quantum electrodynamics are taken by crossing; we
also computed them from their diagrams (for the double boxes see
Refs.~\cite{Smirnov:1999gc,Tausk:1999vh}) and found agreement. Their
infrared poles have the universal form of Ref.~\cite{Catani:1998bh}. The logarithms of
the electron mass factorize~\cite{Mitov:2006xs,Becher:2007cu}. Both cancel in the
asymmetry.

We expand the integrands in the ratios of scales~\cite{AE,Beneke:1997zp}. Each
loop momentum is either hard, of the order of the masses of the weak bosons, or soft, of
the order of $\sqrt s$; a closed loop of a fermion of mass $m_f$ adds regions at that
scale. With two hard momenta the integrals are two-loop vacuum integrals with the masses
$M_W$, $M_Z$, $M_H$ and $m_t$, which reduce to one-loop integrals and a known
function~\cite{Davydychev:1992mt}. Regions with one hard momentum are products of one-loop
integrals. When both momenta are soft the weak boson shrinks to a point. What remains is
the two-loop amplitude of quantum electrodynamics with one contact interaction. Its
integrals are massless four-point functions; we reduce them with
Kira~\cite{Lange:2025fba} to a few master integrals, which are known in closed
form~\cite{Gonsalves:1983nq} or follow from their differential
equations~\cite{Henn:2013pwa}. We checked them numerically~\cite{Liu:2022chg}. 

In closed fermion loops $\gamma_5$ anticommutes. 
A trace with an odd
number of $\gamma_5$ contributes only in fermion triangles. There
$\gamma_5$ is moved to the vertex where the trace starts and written
in the form of Ref.~\cite{Larin:1993tq}. The resulting terms with a
Levi-Civita tensor cancel within a generation of massless
fermions. Starting the trace at another vertex gives the same finite
sum. 
The mixing angle is renormalized by minimal subtraction
with six quarks and converted to $\shZ$ with Eq.~(10.46) of Ref.~\cite{ParticleDataGroup:2026mpi}, the
charge in the Thomson limit and the masses on
shell~\cite{Denner:1991kt}. The muon decay amplitude is needed only in
the hard region, because the photonic corrections of the Fermi theory
are part of the definition of $G_\mu$.

The following checks were made. (i) After the one-loop counterterms,
the two-loop poles are independent of the momenta; their only
dependence on the masses is through the Yukawa coupling of the top
quark. They equal the two-loop counterterm of the mixing angle that
follows from the $\beta$ functions of the gauge
couplings~\cite{Machacek:1983tz,Mihaila:2012fm}, separately for zero,
one and two fermion loops.  
(ii) 
When the mixing angle is subtracted at a scale other than $M_Z$, $\QW$ does not change: the logarithms of the scale in $b_1$, $b_2$, and $b_2^\prime$ cancel the running of the angle through two loops.
(iii) Heavy particles in loops change photon exchange only through the
electric charge, which is the same for both helicities. The hard
regions obey this at leading power in $1/M_W^2$. (iv)
The sum of diagrams is symmetric under the exchange of the two outgoing electrons, as required for identical particles, $\QW(y)=\QW(1-y)$. 
(v) The diagrams with two fermion loops are chains of one-loop
self-energies. Summed to all orders, the chains replace the charge and
the mixing angle by their values at low momentum and multiply the weak
charge by the $\rho$ parameter~\cite{Consoli:1989fg}. The second-order
terms of this product, valid for any $\shZ^2$, equal the sum of our
diagrams.
The ancillary files give the formula. (vi) With separate gauge parameters for the $W$
boson, the $Z$ boson and the photon, the terms of first order in the expansion in each parameter around
the Feynman gauge cancel in $b_2$ and $b_2^\prime$, 
separately for zero, one and two fermion loops.

\emph{Logarithms.}---Let $L=\ln(M_Z^2/s)\approx\Lref$. In powers of $L$ the two-loop
coefficients are
\begin{align}
b_2&=-\frac{11}{6}\,b_1(1/4)\,L+\dots,\nonumber\\
b_2^\prime&=\bigg[\frac12\Big(\frac{11}{6}\Big)^2+\frac{11}{36}\bigg]L^2+\dots,
\label{eq:dl}
\end{align}
where the dots stand for lower powers of $L$. The first term of $b_2^\prime$ comes from the
diagrams without closed fermion loops and adds $+\Bdl\ee$ to $\QW$. It is the
second-order term of the exponential of the one-loop logarithm,
$-\tfrac{11}{6}\tfrac{\alpha}{\pi}(1-4\shZ^2)L$, as the renormalization group (RG) of the theory
below the weak scale requires~\cite{Kollatzsch:2025pnp}; here it follows from the poles of
the diagrams. The second term, $0.02\ee$, comes from the electron loop. In $b_2$ the
logarithm multiplies the complete one-loop coefficient at $\shZ^2=1/4$ with the same
anomalous dimension.

A fermion of mass $m_f\gg\sqrt s$, charge $Q_f$, isospin $T_{3f}$ and $N_f$ colors
contributes $(\alpha/\pi)k_f\ln(m_f^2/M_Z^2)$ to the one-loop term, with
$k_f=\tfrac23N_fQ_f(T_{3f}-2\shZ^2Q_f)$. 
If the fermion is lighter than the $Z$ boson, the two-loop coefficients contain the
squared logarithm
\begin{equation}
-\Big(\frac{\alpha}{\pi}\Big)^2\Big[\frac{11}{12}k_f
+\frac{1-4\shZ^2}{4}N_fQ_f^2(1+2T_{3f})\Big]\ln^2\frac{M_Z^2}{m_f^2}.
\label{eq:cf}
\end{equation}
With $k_f$ at $\shZ^2=1/4$, the first term is part of $b_2$; the rest belongs to $b_2^\prime$.
The first term comes from the one-loop mixing nested inside the integral that gives
Eq.~\eqref{eq:dl}. The second comes from the charge in the $\gamma Z$ box, which runs
with the fermion loop, and from
the triangles in which the fermion loop couples to the $Z$ and to one photon from each electron; the two
cancel for $T_{3f}=-\tfrac12$. In all of them the fermion loop sits inside a photon propagator
whose momentum runs between $\sqrt s$ and $M_Z$, a fact we use for the hadrons.

\emph{Hadrons.}---Light quarks enter through the correlator of the electromagnetic current
with the vector current of each flavor. We give the result for free quarks with masses
$m_q$. For numbers, $\ln(M_Z^2/m_q^2)$ is replaced by the value of the correlator at zero
momentum in minimal subtraction at $M_Z$. We take it from the dispersive analysis of
$e^+e^-$ data of Ref.~\cite{Erler:2017knj} or, for the three light flavors, from lattice
QCD~\cite{Conigli:2025qvh}; the values are listed in the ancillary files.
At two loops the squared logarithms of
Eq.~\eqref{eq:cf} are corrected for the measured momentum dependence of the correlator.
The dependence of the one-loop mixing on the momentum transfer is taken from the
lattice function~\cite{Conigli:2025qvh,Ce:2022eix}. At one loop this reproduces the sum
$\frac12\sum_q k_q\ln(m_q^2/M_Z^2)$ of Ref.~\cite{Czarnecki:1995fw} over the five lighter
quarks, now $-6.75\pm0.02$ (data) or $-6.812\pm0.005$ (lattice) instead of $-6.88\pm0.50$.

QCD corrections to the top-quark loop are included in the $\rho$
parameter~\cite{Djouadi:1987gn,Djouadi:1987di,Kniehl:1989yc,Avdeev:1994db,Chetyrkin:1995ix}
and in the conversion to $\shZ$. Quark triangles enter only through the part that is transverse
in the axial current, so that the pion pole drops out~\cite{Czarnecki:2002nt}; their
hadronic uncertainty is $3\times 10^{-6}$.

\begin{table}[b]
\caption{The weak charge of the electron, Eq.~\eqref{eq:alr}, in units of $10^{-3}$, for
a beam of $11$~GeV, $y=1/2$ ($Q^2=\Qref$~GeV$^2$), $\shZ^2=0.23122$, $m_t=172.6$~GeV and
$M_H=125.1$~GeV, in powers of $\alpha$. The two-loop lines contain the terms of order $\alpha^2$ of the conversion to $\shZ$. Each entry is rounded separately. The lines below $\QW$ are its
uncertainties from the hadronic input and from terms beyond this calculation; the parts
of the latter are added in quadrature. The uncertainties of $m_t$ and $M_H$ change $\QW$
by less than $0.005$.\label{tab:res}}
\begin{ruledtabular}
\begin{tabular}{lrr}
 & data & lattice\\
\hline
$1-4\shZ^2$ & \multicolumn{2}{c}{$\Btree$}\\
one loop, no fermion loop & \multicolumn{2}{c}{$\BoneB$}\\
one loop, fermion loops & $\BoneF$ & $\BoneFL$\\
\quad mixing at $t$ and $u$ & \multicolumn{2}{c}{$\BoneTU$}\\
\quad QCD in the top loop & \multicolumn{2}{c}{$\BoneAS$}\\
two loops, no fermion loop & \multicolumn{2}{c}{$\BtwoB$}\\
two loops, one fermion loop & $\BtwoF$ & $\BtwoFL$\\
two loops, two fermion loops & $\BtwoFF$ & $\BtwoFFL$\\
\quad QCD in the top loop & \multicolumn{2}{c}{$\BtwoAS$}\\
\hline
$\QW$ & $\Bdata$ & $\Blat$\\
\quad hadronic input & $\pm\Edata$ & $\pm\Elat$\\
\quad beyond this calculation & \multicolumn{2}{c}{$\pm\Bpert$}\\
\qquad three loops & \multicolumn{2}{c}{$\pm0.04$}\\
\qquad $\alpha_s$ in hard two-loop terms & \multicolumn{2}{c}{$\pm0.01$}\\
\qquad $G_\mu$ in $d$ dimensions & \multicolumn{2}{c}{$\pm0.005$}\\
\qquad higher orders in $1-4\shZ^2$ & \multicolumn{2}{c}{$\pm0.002$}\\
\end{tabular}
\end{ruledtabular}
\end{table}

\emph{Results.}---Table~\ref{tab:res} lists the contributions to Eq.~\eqref{eq:main} at
$y=1/2$, the center of the MOLLER acceptance, where $A_{LR}=\ALRref\times10^{-9}$ with hadrons
from data. The two-loop terms add $\BtwoSum\ee$, or $\BtwoPct\%$ of the asymmetry, about
two thirds of the experimental uncertainty. The second uncertainty in Eq.~\eqref{eq:main} is our
estimate of the terms beyond this calculation, with its parts in Table~\ref{tab:res}.
Three loops dominate; we estimate them from the leading logarithms.

The number $3.45$  in Eq.~\eqref{eq:angle} is the slope of $\QW$ in $\shZ^2$; it differs from the tree-level value $4$ because the one-loop terms depend  on
the mixing angle (Fig.~\ref{fig:qw}). Per GeV of beam energy $\QW$ changes by
$\dBdE\ee$. The two hadronic inputs differ by $\Bdiff\ee$, four times the uncertainty of
the data and more than ten times that of the lattice result. This is the known tension between
$e^+e^-$ data and lattice QCD in the hadronic vacuum
polarization~\cite{Erler:2024lds,Conigli:2026stl}, which is also at issue in the anomalous
magnetic moment of the muon~\cite{Aliberti:2025beg}. It enters at one loop and is now
the largest uncertainty of the prediction.

Equation~\eqref{eq:alr} with this $\QW$ is the asymmetry from virtual corrections, in
which the infrared singularities cancel. Real photon emission changes the measured
asymmetry by an amount that depends on the acceptance and has to be
simulated~\cite{Banerjee:2021qvi,Kollatzsch:2025pnp}.

\begin{table}[t]
\caption{Comparison with earlier calculations, in units of $10^{-3}$, each at the inputs
and in the scheme of the reference: for Ref.~\cite{Chen:2026qcq}, the \MSbar\  angle $s_0\equiv\sin\hat\theta_W(0)$ at zero momentum; for Ref.~\cite{Du:2019evk}, \MSbar\ with six quarks. For no fermion loop at $s_0^2=1/4$ the value of
Ref.~\cite{Chen:2026qcq} is from its ancillary file. The doubly indented lines split the
difference at the angle of that reference, reference minus this work, into its parts.
The lines ``correlators'' give the change of our result when the threshold quark masses
of Ref.~\cite{Du:2019evk} are replaced by the measured correlators. For the mixing at $t$
and $u$, Ref.~\cite{Erler:2022ckm} gives $0.029$ for all fermions. For the hadronic
effects at two loops it gives a bound; our number there is the shift from the measured
momentum dependence of the correlator.\label{tab:cmp}}
\begin{ruledtabular}
\begin{tabular}{lrr}
 & this work & reference\\
\hline
\multicolumn{3}{l}{no fermion loop~\cite{Chen:2026qcq}}\\
\quad at $s_0^2=1/4$ & $-0.312$ & $-0.308$\\
\quad at the angle of Ref.~\cite{Chen:2026qcq} & $-0.164$ & $-0.134(54)$\\
\qquad squared logarithm & \multicolumn{2}{c}{$0.075$}\\
\qquad difference at $s_0^2=1/4$ & \multicolumn{2}{c}{$0.004$}\\
\qquad terms $\propto1-4s_0^2$ & \multicolumn{2}{c}{$-0.051$}\\
\multicolumn{3}{l}{fermion loops~\cite{Du:2019evk}}\\
\quad one fermion loop & $1.196$ & $1.18$\\
\quad two fermion loops & $-0.226$ & $-0.18$\\
\quad correlators, one loop & $\BThrOne$ & \\
\quad correlators, two loops & $+\BThrTwo$ & \\
\multicolumn{3}{l}{hadronic effects~\cite{Erler:2022ckm}}\\
\quad QCD in the top loop & $-0.05$ & $-0.05$\\
\quad mixing at $t$ and $u$, quarks & $\BoneTU$ & $0.025$\\
\quad two loops, one fermion loop & $+\BmeasF$ & $\pm0.06$\\
\multicolumn{3}{l}{all terms~\cite{Chen:2026qcq}}\\
\quad $\QW$, hadrons from data & $\Bdata$ & $45.494$\\
\end{tabular}
\end{ruledtabular}
\end{table}

\emph{Comparison with earlier results.}---Table~\ref{tab:cmp} compares
our results with earlier calculations.  In comparing with
Ref.~\cite{Chen:2026qcq} the only difference we find is the squared
logarithm $L^2$: its coefficient there is twice the first term of
$b_2^\prime$ in Eq.~\eqref{eq:dl}, and in the RG prediction
\cite{Kollatzsch:2025pnp}.  The coefficient of $L^2$ is suppressed by
$1-4\shZ^2$ so the effect is small, $0.075\ee$, and is offset by
single logarithms and constants proportional to $1-4\shZ^2$, neglected
in Ref.~\cite{Chen:2026qcq}, so that the totals agree within its
uncertainty.

Two differences from Ref.~\cite{Du:2019evk} stand out. It approximates
the hadronic vacuum polarization by threshold quark masses; the
measured correlators move the one-loop term by three times the
experimental goal and the two-loop terms by a tenth of it. 

The term with two fermion loops in Ref.~\cite{Du:2019evk} is higher
than ours by $0.046\ee$, an amount that resembles the product of the
top-quark term of the $\rho$ parameter with the running of $\alpha$
from zero momentum to $M_Z$. The product cancels in our expression,
between the normalization to $G_\mu$ and the mixing angle at low
energy; 
the difference may come from the treatment of this product in Ref.~\cite{Du:2019evk}.

Reference~\cite{Aleksejevs:2011de} found that the products of one-loop
amplitudes, in the on-shell scheme, raise the asymmetry by up to $4\%$
of its tree-level value. This large correction is misleading because the product of one-loop amplitudes, separated from two-loop diagrams, depends on the renormalization scheme. In fact, 
all two-loop terms together raise the asymmetry by only 
$0.9\%$ of its tree-level value ($\BtwoPct\%$ of the asymmetry).

\emph{Conclusions.}---The two-loop electroweak corrections to the asymmetry of M\o ller
scattering are now known for all classes of diagrams and beyond the limit
$\sin^2\theta_W=1/4$. They contribute $\BtwoPct\%$ of the weak charge of the electron;
the uncertainty from terms beyond this calculation, $\Bpert\ee$, is $4\%$ of the goal of
the MOLLER experiment. The prediction is limited by the hadronic vacuum polarization at
one loop: $e^+e^-$ data and lattice QCD give values of $\QW$ that differ by
$\BdiffPct\%$, about a quarter of the goal. This difference has to be resolved, or
carried as an uncertainty, when the measured asymmetry is compared with the Standard
Model.

The differences we found with earlier studies are a few percent of the planned precision of MOLLER.
We thus conclude that the Standard Model prediction stands on solid ground. The next turn belongs to the experimenters.

The ancillary files~\cite{Czarnecki:2026anc} also contain the hadronic input of
Table~\ref{tab:res}, a catalogue of the diagrams and the program, which gives the lines
of Fig.~\ref{fig:qw} and evaluates Eq.~\eqref{eq:exp} for other values of the mixing
angle, the beam energy and $y$.

\emph{Data availability.}---The data and the program that support the findings of this
article are publicly available~\cite{Czarnecki:2026anc}.

\begin{acknowledgments}
  This work made substantive use of an AI system: the models Claude
  Opus~5.5 and Claude Fable~5.1 (Anthropic).  Following instructions
  of the author, it wrote and ran the programs for computer algebra
  and for the numerical evaluation, which are organized like earlier
  programs of the author, carried out the algebra of the calculation,
  took part in the survey of the literature, compared our results with
  it, and prepared a first draft of the text and of the figure. The
  author chose the method, the approximations and the input
  parameters, proposed the expansion around the Feynman gauge with
  which the dependence on the gauge parameters is tested, proposed to
  test the poles of the region with two soft momenta against the
  renormalization group of the effective theory below the weak scale,
  surveyed part of the literature, and rewrote the text. The results
  were verified by the checks (i) to (vi) listed above, by the
  reproduction of the one-loop result of Ref.~\cite{Czarnecki:1995fw},
  and by the comparisons with
  Refs.~\cite{Du:2019evk,Chen:2026qcq}. The author takes full
  responsibility for the content of this paper.

This research was supported by the Natural Sciences and Engineering Research Council of Canada (NSERC). 
\end{acknowledgments}


\begin{thebibliography}{10}
\providecommand{\url}[1]{\texttt{#1}}
\providecommand{\urlprefix}{URL }
\providecommand{\eprint}[2][]{\url{#2}}

\bibitem{Derman:1979zc}
E.~Derman and W.~J. Marciano, \emph{{Parity Violating Asymmetries in Polarized
  Electron Scattering}}, Annals Phys. \textbf{121}, 147 (1979).

\bibitem{Kumar:2013yoa}
K.~S. Kumar, S.~Mantry, W.~J. Marciano, and P.~A. Souder, \emph{{Low-Energy
  Measurements of the Weak Mixing Angle}}, Annu. Rev. Nucl. Part. Sci.
  \textbf{63}, 237--267 (2013), \eprint{1302.6263}.

\bibitem{Czarnecki:1995fw}
A.~Czarnecki and W.~J. Marciano, \emph{{Electroweak radiative corrections to
  polarized M{\o}ller scattering asymmetries}}, Phys. Rev. D \textbf{53},
  1066--1072 (1996), \eprint{hep-ph/9507420}.

\bibitem{SLACE158:2005uay}
P.~L. Anthony et~al., \emph{{Precision Measurement of the Weak Mixing Angle in
  M{\o}ller Scattering}}, Phys. Rev. Lett. \textbf{95}, 081601 (2005),
  \eprint{hep-ex/0504049}.

\bibitem{Czarnecki:2005pe}
A.~Czarnecki and W.~J. Marciano, \emph{Electrons are not ambidextrous}, Nature
  \textbf{435}, 437--438 (2005).

\bibitem{MOLLER:2014iki}
J.~Benesch et~al., \emph{{The MOLLER Experiment: An Ultra-Precise Measurement
  of the Weak Mixing Angle Using M{\o}ller Scattering}} (2014),
  \eprint{1411.4088}.

\bibitem{ALEPH:2005ab}
S.~Schael et~al., \emph{{Precision electroweak measurements on the $Z$
  resonance}}, Phys. Rep. \textbf{427}, 257--454 (2006),
  \eprint{hep-ex/0509008}.

\bibitem{Marciano:1999ih}
W.~J. Marciano, \emph{{Fermi Constants and ``New Physics''}}, Phys. Rev. D
  \textbf{60}, 093006 (1999), \eprint{hep-ph/9903451}.

\bibitem{ParticleDataGroup:2026mpi}
F.~Takahashi et~al., \emph{{Review of Particle Physics}}, Int. J. Mod. Phys. A
  \textbf{41}, 2630011 (2026).

\bibitem{Marciano:1990dp}
W.~J. Marciano and J.~L. Rosner, \emph{{Atomic parity violation as a probe of
  new physics}}, Phys. Rev. Lett. \textbf{65}, 2963--2966 (1990), [Erratum:
  Phys. Rev. Lett. \textbf{68}, 898 (1992)].

\bibitem{Fanchiotti:1992tu}
S.~Fanchiotti, B.~A. Kniehl, and A.~Sirlin, \emph{{Incorporation of QCD effects
  in basic corrections of the electroweak theory}}, Phys. Rev. D \textbf{48},
  307--331 (1993), \eprint{hep-ph/9212285}.

\bibitem{Gambino:1993dd}
P.~Gambino and A.~Sirlin, \emph{{Relation between $\sin^2\hat\theta_W(m_Z)$ and
  $\sin^2\theta_{\rm eff}^{\rm lept}$}}, Phys. Rev. D \textbf{49}, R1160--R1162
  (1994), \eprint{hep-ph/9309326}.

\bibitem{Czarnecki:2026anc}
A.~Czarnecki, \emph{{Ancillary files of ``Parity-violating M{\o}ller
  scattering: complete two-loop corrections''}}, Zenodo (2026),
  doi:10.5281/zenodo.23681692.

\bibitem{Nogueira:1991ex}
P.~Nogueira, \emph{{Automatic Feynman graph generation}}, J. Comput. Phys.
  \textbf{105}, 279--289 (1993).

\bibitem{Vermaseren:2000nd}
J.~A.~M. Vermaseren, \emph{{New features of FORM}} (2000), \eprint{math-ph/0010025}.

\bibitem{Davies:2026cci}
J.~Davies, T.~Kaneko, C.~Marinissen, T.~Ueda, and J.~A.~M. Vermaseren,
  \emph{{FORM Version 5.0}} (2026), \eprint{2601.19982}.

\bibitem{Glover:2004si}
E.~W.~N. Glover, \emph{{Two loop QCD helicity amplitudes for massless quark
  quark scattering}}, J. High Energy Phys. \textbf{04}, 021 (2004),
  \eprint{hep-ph/0401119}.

\bibitem{Smirnov:1999gc}
V.~A. Smirnov, \emph{{Analytical result for dimensionally regularized massless
  on-shell double box}}, Phys. Lett. B \textbf{460}, 397--404 (1999),
  \eprint{hep-ph/9905323}.

\bibitem{Tausk:1999vh}
J.~B. Tausk, \emph{{Non-planar massless two-loop Feynman diagrams with four
  on-shell legs}}, Phys. Lett. B \textbf{469}, 225--234 (1999),
  \eprint{hep-ph/9909506}.

\bibitem{Catani:1998bh}
S.~Catani, \emph{{The singular behaviour of QCD amplitudes at two-loop order}},
  Phys. Lett. B \textbf{427}, 161--171 (1998), \eprint{hep-ph/9802439}.

\bibitem{Mitov:2006xs}
A.~Mitov and S.~Moch, \emph{{The singular behavior of massive QCD amplitudes}},
  J. High Energy Phys. \textbf{05}, 001 (2007), \eprint{hep-ph/0612149}.

\bibitem{Becher:2007cu}
T.~Becher and K.~Melnikov, \emph{{Two-loop QED corrections to Bhabha
  scattering}}, J. High Energy Phys. \textbf{06}, 084 (2007),
  \eprint{0704.3582}.

\bibitem{AE}
For reviews and further references see, for example,\\ F.~V.~Tkachov,
  \emph{Theory of asymptotic operation. A summary of basic principles}, Sov.\
  J.\ Part.\ Nucl.\ {\bf 25}, 649 (1994), arXiv:hep-ph/9701272; \\
  V.~A.~Smirnov, {\em Applied asymptotic expansions in momenta and masses},
  Springer Verlag (Berlin), 2002; \\ K. G. Chetyrkin, \emph{Combinatorics of
  $\mathbf{R}$, $\mathbf{R^{-1}}$, and $\mathbf{R^*}$-operations and asymptotic
  expansions of Feynman integrals in the limit of large momenta and masses},
  arXiv:1701.08627; \\ G. B. Pivovarov and F. V. Tkachov, \emph{Euclidean
  Asymptotic Expansions of Green Functions of Quantum Fields: II Combinatorics
  of the AS Operation}, Int. J. Mod. Phys. {\bf A8} (1993) 2241,
  arXiv:hep-ph/9612287.

\bibitem{Beneke:1997zp}
M.~Beneke and V.~A. Smirnov, \emph{{Asymptotic expansion of Feynman integrals
  near threshold}}, Nucl. Phys. B \textbf{522}, 321--344 (1998),
  \eprint{hep-ph/9711391}.

\bibitem{Davydychev:1992mt}
A.~I. Davydychev and J.~B. Tausk, \emph{{Two-loop self-energy diagrams with
  different masses and the momentum expansion}}, Nucl. Phys. B \textbf{397},
  123--142 (1993).

\bibitem{Lange:2025fba}
F.~Lange, J.~Usovitsch, and Z.~Wu, \emph{{Kira 3: integral reduction with
  efficient seeding and optimized equation selection}}, Comput. Phys. Commun.
  \textbf{322}, 109999 (2026), \eprint{2505.20197}.

\bibitem{Gonsalves:1983nq}
R.~J. Gonsalves, \emph{{Dimensionally regularized two-loop on-shell quark form
  factor}}, Phys. Rev. D \textbf{28}, 1542 (1983).

\bibitem{Henn:2013pwa}
J.~M. Henn, \emph{{Multiloop integrals in dimensional regularization made
  simple}}, Phys. Rev. Lett. \textbf{110}, 251601 (2013), \eprint{1304.1806}.

\bibitem{Liu:2022chg}
X.~Liu and Y.-Q. Ma, \emph{{AMFlow: A Mathematica package for Feynman integrals
  computation via auxiliary mass flow}}, Comput. Phys. Commun. \textbf{283},
  108565 (2023), \eprint{2201.11669}.

\bibitem{Larin:1993tq}
S.~A. Larin, \emph{{The Renormalization of the axial anomaly in dimensional
  regularization}}, Phys. Lett. B \textbf{303}, 113--118 (1993),
  \eprint{hep-ph/9302240}.

\bibitem{Denner:1991kt}
A.~Denner, \emph{{Techniques for calculation of electroweak radiative
  corrections at the one loop level and results for W physics at LEP-200}},
  Fortsch. Phys. \textbf{41}, 307--420 (1993), \eprint{0709.1075}.

\bibitem{Machacek:1983tz}
M.~E. Machacek and M.~T. Vaughn, \emph{{Two-loop renormalization group
  equations in a general quantum field theory: (I). Wave function
  renormalization}}, Nucl. Phys. B \textbf{222}, 83--103 (1983).

\bibitem{Mihaila:2012fm}
L.~N. Mihaila, J.~Salomon, and M.~Steinhauser, \emph{{Gauge coupling beta
  functions in the standard model to three loops}}, Phys. Rev. Lett.
  \textbf{108}, 151602 (2012), \eprint{1201.5868}.

\bibitem{Consoli:1989fg}
M.~Consoli, W.~Hollik, and F.~Jegerlehner, \emph{{The Effect of the Top Quark
  on the M(W)-M(Z) Interdependence and Possible Decoupling of Heavy Fermions
  from Low-Energy Physics}}, Phys. Lett. B \textbf{227}, 167--170 (1989).

\bibitem{Kollatzsch:2025pnp}
S.~Kollatzsch, D.~Moreno, D.~Radic, and A.~Signer, \emph{{Parity violation in
  M{\o}ller scattering within low-energy effective field theory}}, J. High
  Energy Phys. \textbf{09}, 196 (2025), \eprint{2507.17652}.

\bibitem{Erler:2017knj}
J.~Erler and R.~Ferro-Hern{\'a}ndez, \emph{{Weak mixing angle in the Thomson
  limit}}, J. High Energy Phys. \textbf{03}, 196 (2018), \eprint{1712.09146}.

\bibitem{Conigli:2025qvh}
A.~Conigli, D.~Djukanovic, G.~von Hippel, S.~Kuberski, H.~B. Meyer, K.~Miura,
  K.~Ottnad, A.~Risch, and H.~Wittig, \emph{{Precision lattice calculation of
  the hadronic contribution to the running of the electroweak gauge
  couplings}}, Phys. Rev. D \textbf{114}, 034508 (2026), \eprint{2511.01623}.

\bibitem{Ce:2022eix}
M.~C{\`e}, A.~G{\'e}rardin, G.~von Hippel, H.~B. Meyer, K.~Miura, K.~Ottnad,
  A.~Risch, T.~San~Jos{\'e}, J.~Wilhelm, and H.~Wittig, \emph{{The hadronic
  running of the electromagnetic coupling and the electroweak mixing angle from
  lattice QCD}}, J. High Energy Phys. \textbf{08}, 220 (2022),
  \eprint{2203.08676}.

\bibitem{Djouadi:1987gn}
A.~Djouadi and C.~Verzegnassi, \emph{{Virtual very heavy top effects in LEP/SLC
  precision measurements}}, Phys. Lett. B \textbf{195}, 265--271 (1987).

\bibitem{Djouadi:1987di}
A.~Djouadi, \emph{{${\cal O}(\alpha\alpha_s)$ vacuum polarization functions of
  the standard-model gauge bosons}}, Nuovo Cimento A \textbf{100}, 357 (1988).

\bibitem{Kniehl:1989yc}
B.~A. Kniehl, \emph{{Two-loop corrections to the vacuum polarizations in
  perturbative QCD}}, Nucl. Phys. B \textbf{347}, 86--104 (1990).

\bibitem{Avdeev:1994db}
L.~Avdeev, J.~Fleischer, S.~Mikhailov, and O.~Tarasov, \emph{{${\cal
  O}(\alpha\alpha_s^2)$ correction to the electroweak $\rho$ parameter}}, Phys.
  Lett. B \textbf{336}, 560--566 (1994), [Erratum: Phys. Lett. B \textbf{349},
  597 (1995)], \eprint{hep-ph/9406363}.

\bibitem{Chetyrkin:1995ix}
K.~G. Chetyrkin, J.~H. K{\"u}hn, and M.~Steinhauser, \emph{{Corrections of
  order ${\cal O}(G_F M_t^2 \alpha_s^2)$ to the $\rho$ parameter}}, Phys. Lett.
  B \textbf{351}, 331--338 (1995), \eprint{hep-ph/9502291}.

\bibitem{Czarnecki:2002nt}
A.~Czarnecki, W.~J. Marciano, and A.~Vainshtein, \emph{{Refinements in
  electroweak contributions to the muon anomalous magnetic moment}}, Phys. Rev.
  D \textbf{67}, 073006 (2003), [Erratum: Phys. Rev. D \textbf{73}, 119901
  (2006)], \eprint{hep-ph/0212229}.

\bibitem{Erler:2024lds}
J.~Erler, R.~Ferro-Hern{\'a}ndez, and S.~Kuberski, \emph{{Theory-Driven
  Evolution of the Weak Mixing Angle}}, Phys. Rev. Lett. \textbf{133}, 171801
  (2024), \eprint{2406.16691}.

\bibitem{Conigli:2026stl}
A.~Conigli, D.~Djukanovic, G.~von Hippel, S.~Kuberski, H.~B. Meyer, K.~Miura,
  K.~Ottnad, A.~Risch, and H.~Wittig, \emph{{Running of the Electroweak Gauge
  Couplings from First Principles}}, Phys. Rev. Lett. \textbf{137}, 071901
  (2026), \eprint{2607.03370}.

\bibitem{Aliberti:2025beg}
R.~Aliberti et~al., \emph{{The anomalous magnetic moment of the muon in the
  Standard Model: an update}}, Phys. Rep. \textbf{1143}, 1--158 (2025),
  \eprint{2505.21476}.

\bibitem{Banerjee:2021qvi}
P.~Banerjee, T.~Engel, N.~Schalch, A.~Signer, and Y.~Ulrich, \emph{{M{\o}ller
  scattering at NNLO}}, Phys. Rev. D \textbf{105}, L031904 (2022),
  \eprint{2107.12311}.

\bibitem{Chen:2026qcq}
L.~Chen, J.~Erler, A.~Freitas, and J.~Kwak, \emph{{Bosonic Contribution to
  Parity-Violating M{\o}ller Scattering at NNLO}} (2026), \eprint{2610.12287}.

\bibitem{Du:2019evk}
Y.~Du, A.~Freitas, H.~H. Patel, and M.~J. Ramsey-Musolf,
  \emph{{Parity-Violating M{\o}ller Scattering at Next-to-Next-to-Leading
  Order: Closed Fermion Loops}}, Phys. Rev. Lett. \textbf{126}, 131801 (2021),
  \eprint{1912.08220}.

\bibitem{Erler:2022ckm}
J.~Erler, R.~Ferro-Hern{\'a}ndez, and A.~Freitas, \emph{{Hadronic effects in
  M{\o}ller scattering at NNLO}}, J. High Energy Phys. \textbf{08}, 183 (2022),
  \eprint{2202.11976}.

\bibitem{Aleksejevs:2011de}
A.~Aleksejevs, S.~Barkanova, Y.~Kolomensky, E.~Kuraev, and V.~Zykunov,
  \emph{{Quadratic electroweak corrections for polarized M{\o}ller
  scattering}}, Phys. Rev. D \textbf{85}, 013007 (2012), \eprint{1110.1750}.

\end{thebibliography}
\end{document}